# Magnetic detection under high pressures using designed silicon vacancy centers in silicon carbide

Jun-Feng Wang[1,3,4, †], Lin Liu[2,1, †], Xiao-Di Liu[2, †, *], Qiang Li[1,4], Jin-Ming Cui[1,4,7], Di-Fan Zhou[5], Ji-Yang Zhou[1,4], Yu Wei[6], Hai-An Xu[2], Wan Xu[2], Wu-Xi Lin[1,4], Jin-Wei Yan[2], Zhen-Xuan He[1,4], Zheng-Hao Liu[1, 4], Zhi-He Hao[1, 4], Hai-Ou Li[1, 4, 7], Wen Liu[6], Jin-Shi Xu[1, 4, 7*], Eugene Gregoryanz[2, *], Chuan-Feng Li[1, 4, 7*] and Guang-Can Guo[1, 4, 7]

[1]*CAS Key Laboratory of Quantum Information, University of Science and Technology of China, Hefei 230026, China*

[2]*Key Laboratory of Materials Physics, Institute of Solid State Physics, HFIPS, Chinese Academy of Sciences, Hefei 230031, China*

[3]*College of Physics, Sichuan University, Chengdu, Sichuan 610065, China*

[4]*CAS Center for Excellence in Quantum Information and Quantum Physics, University of Science and Technology of China, Hefei 230026, China*

[5]*Physics Department, Shanghai Key Laboratory of High Temperature Superconductors, Shanghai University, Shanghai 200444, China*

[6]*Center for Micro- and Nanoscale Research and Fabrication, University of Science and Technology of China, Hefei 230027, China*

[7]*Hefei National Laboratory, University of Science and Technology of China, Hefei 230088, China*

[†]These authors contributed equally to the work.

[*]Corresponding author: xiaodi@issp.ac.cn, jsxu@ustc.edu.cn, e.gregoryanz@ed.ac.uk, cfli@ustc.edu.cn

## Abstract

Pressure-induced magnetic phase transition is attracting interest due to its ability to detect superconducting behaviour at high pressures in diamond anvil cells. However, detection of the local sample magnetic properties is a great challenge due to the small sample chamber volume. Recently, optically detected magnetic resonance (ODMR) of nitrogen vacancy (NV) centers in diamond have been used for in-situ pressure-induced phase transition detection. However, owing to their four orientation axes and temperature-dependent zero-field-splitting, interpreting the observed ODMR spectra of NV centers remain challenging. Here, we study the optical and spin properties of implanted silicon vacancy defects in 4H-SiC, which is single-axis and temperature-independent zero-field-splitting. Using this technique, we observe the magnetic phase transition of $Nd_2Fe_{14}B$ at about 7 GPa and map the critical temperature-pressure phase diagram of the superconductor $YBa_2Cu_3O_{6.6}$. These results highlight the potential of silicon vacancy-based quantum sensors for in-situ magnetic detection at high pressures.

The ability of pressure to alter the electronic, magnetic and structural properties of matter is a vital feature widely used in fundamental and applied sciences studies [1-8]. High-pressure techniques have been applied in many fields, including physics, material sciences, geophysics and chemistry, leading to many unusual and important phenomena that have been observed under pressure[1-8]. In particular, the claims of pressure-induced high critical temperature ($T_c$) superconductivity have drawn serious attention and excitement in recent years[6-8]. For example, lanthanum hydride has been inferred to be a superconductor with a critical temperature of ~250 K at around 170 GPa[7,8]. One of the great challenges of high-pressure research is the measurement of magnetic properties and their evolution. However, conventional methods such as using superconducting quantum interference devices (SQUIDs) or AC susceptibility cannot directly detect weak magnetic signals of micrometre-sized samples in diamond anvil cell (DAC)[9-14]. Therefore, it is important to explore new methods for magnetic detection.

The high sensitivity and high resolution of *in-situ* magnetic detections in DAC chamber were achieved by utilizing nitrogen-vacancy (NV) centres[10-12]. Diamond NV centres are versatile solid-state spin quantum sensors that have been used to detect a wide variety of physical parameters, such as magnetic and electric fields, temperature, strain, spins, pressure and electrical currents[9-19]. The zero-field-splitting (ZFS) parameter $D$ of the NV centre ground spin state was shown to increase linearly with pressure with a slope of 14.6 MHz/GPa up to 60 GPa[9]. On this basis, an *in-situ* magnetic detection method based on NV centres through optically detected magnetic resonance (ODMR) technologies has recently been developed at high pressure[10-13]. Micron-sized diamond particles with ensemble NV centres have been placed inside the DAC chamber to measure the $T_c$-pressure phase diagram of a superconductor[10] and detect the pressure-induced magnetic phase transition of a magnet[13]. The shallow implanted NV centres on the surface of the diamond were also used to probe the magnetization of Fe particles and the Meissner effect of a superconductor and construct the full stress tensor on the culet surface[11,12].

Defects in silicon carbide (SiC) could also be utilized to measure magnetic properties under extreme conditions. SiC is a widely used semiconductor due to its unique properties, such as mature inch-scale growth and micro/nanofabrication[20-22]. Also, several spin qubits and bright single-photon emitters in SiC have attracted great attention in the quantum community[20-32]. In particular, the silicon vacancy defect at the hexagonal lattice site with a negative charge ($V_{Si}$) has been extensively used in spin-photon interfaces[23], quantum photonics[26], quantum information processing[22], and quantum sensing, such as magnetic fields[30] and temperatures[31,32] due to its outstanding properties. Its spin state is an $S = 3/2$ spin quartet, and the ground state ZFS parameter $D$ is ~70 MHz[22,23]. It only has one axis (along the c-axis of the 4H-SiC), and the corresponding ODMR spectrum has two resonant peaks under an external magnetic field, which is convenient to readout the resonant frequencies and enhancing the scalability in SiC devices[22]. Moreover, the ZFS parameter $D$ is also almost temperature independent from 20 K to 500 K at ambient pressure, which is beneficial to temperature–pressure research[31,32]. However, most of the previous investigations on the $V_{Si}$ defect were performed under ambient pressure[22-32]. The study of the optical and spin properties under high pressure is important for $V_{Si}$ defect-based quantum sensing at extreme conditions. In comparison with traditional high-pressure magnetometry techniques, the spatial resolution of $V_{Si}$ defect detection is only around a few microns.

Here, we investigate and characterize the optical and spin properties of the implanted silicon vacancy defects at the culets of the 4H-SiC anvil, which exhibit single-axis and temperature-independent zero-field splitting. The experimental results show that the photoluminescence (PL) spectrum blueshifts and the ZFS parameter $D$ increases with pressure at a rate of 0.31 MHz/GPa. We probed the pressure-induced magnetic phase transition of a $Nd_2Fe_{14}B$ magnet at around 7 GPa at room temperature. Finally, the Meissner effect of a $YBa_2Cu_3O_{6.6}$ superconductor at different pressures was observed, yielding its $T_c$-$P$ phase diagram. These experiments demonstrate the feasibility of using $V_{Si}$ defects in SiC as novel quantum sensors and open up applications to study superconducting phenomena under extreme conditions.

**Optical properties of silicon vacancies under high pressure**

The experimental configuration used in our experiments is shown in Fig 1a (For the experimental details, see the Method section and Supplementary text 1). First, we describe the optical and spin properties of the $V_{Si}$ defects at high pressures. The energy levels of the defects at high pressures are shown in Fig. 1b. The 720 nm laser pumps the electrons from the ground state to the phonon sideband, and the ZPL at ambient pressure is 916 nm. Both the ZPL and the ground spin state ZFS parameter $D$ change under high pressure. The room temperature PL spectra of the defects at three different compressions are shown in Fig. 1c. The PL spectra of the $V_{Si}$ defects are blue-shifted with pressure. We then investigate the mean counts of the $V_{Si}$ defects as a function of compression. The counts increase as the pressure increases from ambient pressure to 8 GPa caused by the higher detection efficiency at shorter wavelengths of the single-photon counting module (see Fig. 1d). Then, the counts decrease as the pressure increases to approximately 25 GPa (see Supplementary text 1 for more details). At the same time, we observe the decrease in the ODMR contrast with increasing pressure (see Fig. 2a). We speculate that the decrease in the photon counts and ODMR contrast are both related to and driven by the lattice distortion of the

4H-SiC, caused by the inhomogeneity and deviation of compression at high pressures. The altered probability density and the electronic structure of the silicon vacancy due to compression may also contribute to the decrease of the photon counts and ODMR contrast[9,33-35].

**Spin properties of silicon vacancies under high pressure**

We then study the $V_{Si}$ defect spin properties at high pressures. The ODMR spectra at zero external magnetic field are shown in Fig. 2a. The initial zero-pressure ODMR peak of 72.4 ± 0.4 MHz may be due to the strain during the preparation of the SiC anvil, and the effect has been observed before[36-38]. The resonant frequency shifts to higher frequencies as the pressure increases, in line with the ODMR signal of NV centres in diamond[9-14,33-35]. The local structural distortions and the decreasing distance between $V_{Si}$ spin in the macroscopic compression in the SiC crystal drive the resonant frequency shifts to higher values as the pressure increases[9,33-35]. As shown in Fig. 2b, the mean ZFS parameter $D$ increases linearly with the pressure with a coefficient of 0.31 ± 0.01 MHz/GPa, which is considerably smaller than 14.6 MHz/GPa of the NV centres in diamond[9,13,14]. The smaller slope is beneficial for directly observing the shift of the ODMR signal over a large pressure range. The reason for the small slope is due to the degeneracy of half-integer $V_{Si}$ defects (S=3/2), which makes it rather insensitive to strain fluctuations[39].

Through the coherent control of $V_{Si}$ defects, one could detect the noise spectroscopy of the magnetic materials[12]. Fig. 2c shows the measurement of the Rabi oscillation at ambient pressure using a standard pulse sequence[20,32]. Inferred from the fitting, the Rabi frequency is 9 MHz. Figures 2d and 2e present the spin echo measurement of $V_{Si}$ defects at ambient pressure and at 15.1 GPa, yielding the coherence times $T_2$ of 7.8 ± 0.9 μs and 7.3 ± 0.7 μs, respectively. Both values are consistent with previous results[32]. The coherence time $T_2$ as a function of pressure up to 25 GPa is summarized in Fig. 2f. The coherence time remains invariable up to 25 GPa, which is similar to that of NV centres in diamond[13].

**Magnetic detection using silicon vacancies under high pressures**

The SiC anvils with $V_{Si}$ defects could be used to study magnetic and superconducting properties of materials under compression. By utilising the ODMR spectrum, we have studied the pressure-induced magnetic phase transitions of a common magnet $Nd_2Fe_{14}B$. A small piece of $Nd_2Fe_{14}B$ sample is placed on the surface of the culets. The PL images of the implanted shallow $V_{Si}$ defects and $Nd_2Fe_{14}B$ sample on the culet surface are presented in Fig. 3a. To efficiently detect the magnetic field of the sample, a location close to the sample (black dashed line region) is chosen as the detection position, which is denoted with a black cross. As a comparison, a remote location (denoted with a red cross) is the reference position. In the experiment, we apply a c-axis (perpendicular to the culet) magnetic field $B_c$ with a strength of 198 Gauss. Three schematics of local magnetic field vectors at the detected position under different pressures are shown in Fig. 3b. $B_c$, $B_{NdFeB}$ and $B_{tot}$ represent the c-axis external magnetic field, magnetic field of the $Nd_2Fe_{14}B$ sample and total magnetic field on the $V_{Si}$ defects, respectively. Standard lock-in technology is used to detect the ODMR signals[20,32]. The integration time for one frequency is around 5 seconds, and the total measurement time is ~580 seconds. The representative ODMR signals at the detected positions and reference during the compression process are presented in Fig. 3c. The ODMR signals at the reference position reflecting the strength of the external magnetic field ($B_c$) are also measured at each pressure. The ODMR resonant frequencies at the detected position do not change up to 5.1 GPa, but then they abruptly shift to a higher frequency at 6.7 GPa. Since both $B_{tot}$ and $B_c$ could be deduced from the measured ODMR spectra at each pressure, we calculate the magnetic field of the $Nd_2Fe_{14}B$ sample as $|B_{tot}-B_c|$ and plot it in Fig. 3d. The magnetic field of the sample during the compression (blue squares) and decompression (red dots) processes are shown in Fig. 3d. The sample magnetic field, as seen with the ODMR frequencies, stays unchanged as the pressure increases to approximately 6 GPa, but then it has a sharp reduction at around 7 GPa. This phenomenon demonstrates that the $Nd_2Fe_{14}B$ sample reversibly changes from a ferromagnetic phase to a paramagnetic phase at ~7 GPa, in good agreement with the literature[13,40].

Recently, extreme conditions have been applied to synthesize and study novel superconducting materials claiming the critical temperatures well above 200 K[6-8]. As a proof-of-concept experiment, we detected the superconducting phase transition of the well-known superconductor $YBa_2Cu_3O_x$[41,42] at different pressures and low temperatures using our SiC anvils with $V_{Si}$ defects. $YBa_2Cu_3O_x$ is a type II high-$T_c$ superconductor with different concentrations of oxygen x. $YBa_2Cu_3O_{6.6}$ was chosen due to its high Tc and dramatic Tc-pressure curve[41]. $YBa_2Cu_3O_{6.6}$ sample was synthesized in-house by conventional heat treatment methods (see Supplementary text 2 for more details). The confocal scanning image of $V_{Si}$ defects and the $YBa_2Cu_3O_{6.6}$ sample on the culet is marked in Fig. 4a: the sample (black dashed line region) and the detected position (black cross). To measure the superconductor magnetic moment, we first cool the superconductor below its $T_c$ in a zero magnetic field, and then a small c-axis magnetic field (7.7 Gauss) is applied to generate a Zeeman splitting of the $V_{Si}$ defects[43-45]. The ODMR measurements are performed as the temperature increases. The raw ODMR spectra versus temperature at one pressure point (9 GPa) are shown in Fig. 4b. At 9 GPa, the splitting goes through a sudden step-like change at 95 K (Fig. 4c). This is the manifestation of the Meissner effect and the indication that the sample entered the diamagnetic state associated with the superconductivity of the sample. The red line represents the fitting of the data points using a sigmoid function: *S(T)=a+b/(1+exp[-(T-T$_c$)/δT$_c$])*, where *a, b, δT$_c$* are fitting parameters, and $T_c$ is the critical temperature[43,45]. The fitted critical temperature $T_c$ at 9 GPa yields 95.2 ± 0.2 K, which is in excellent agreement with the previous results[41].

We further investigate the critical $T_c$ at different pressures. Fig. 4d shows the measured ODMR splitting as a function of temperature at different pressures. The critical temperature $T_c$ linearly increases with pressure, changing slope at around 12 GPa but continuing to increase (see Fig. 4e). The shadowed and transparent areas represent the superconducting and normal states for $YBa_2Cu_3O_{6.6}$, respectively. Our mapping of the $T_c$ phase diagram is in excellent agreement with the previous data obtained by the AC susceptibility methods in the DAC[41]. The pressure dependence of $T_c$ is because high pressure leading to a change in the charge carrier concentration in the $CuO_2$ planes within the unit cell[42].

**Outlook**

The $V_{Si}$-based *in-situ* magnetic detection technologies could open up several immediate research possibilities in materials science. First, using the higher NA objective, better detectors[11] and optimized samples, both the sensitivity and spatial resolution can be improved several times. The ideal spatial resolution could extend to approximately 1 μm. Since the size of the vortex/domains is approximately micrometre scale, it can be used to detect the magnetic vortices/domains walls of ferromagnetic materials[11,46,47], magnetic 2D-materials[48,49] and geochemistry at high pressure. Second, we could apply the magnetic sensor to investigate the $T_c$-P phase diagram, lower critical magnetic field and London penetration depth of new types of superconductors, such as kagome superconductors at high pressures[44,50]. The 4H-SiC micrometre particles with $V_{Si}$ defects[51] and other types of spin qubits, including divacancies[20,21,52], NV centres[27-29] and even transition metal ions[53] in 4H, 6H and 3C polytypes of SiC, may also be applied to local magnetic detection at high pressure. Some types of novel spin readout technologies, such as photocurrent-detected magnetic resonance[37,54] and anti-Stokes excited ODMR technology[32] can also be used for $V_{Si}$ defect-based magnetic sensing under high pressure. The experiments form a framework for using SiC $V_{Si}$ defects in local *in-situ* magnetic detection under high pressure.

In conclusion, we realize *in-situ* magnetic detection of magnetic materials using an implanted $V_{Si}$ defect ensemble in SiC-based anvil cells under high pressures. By studying the optical and spin properties of the implanted $V_{Si}$ defects, the experiments show that the PL spectrum has a blueshift and the mean counts decrease under high pressure. At the same time, the ZFS parameter D increases with pressure with a small coefficient of 0.31 MHz/GPa, which is much less than that of the NV centres in diamond. Moreover, the spin coherence time remains invariable with pressure, which is vital to probe noise spectroscopy of magnetic materials at high pressure without a direct magnetic signal. Based on these results, the pressure-induced magnetic phase transitions of the magnet $Nd_2Fe_{14}B$ sample are detected in the range of 6-10 GPa using the ODMR methods at room temperature. Finally, we map the superconductor $YBa_2Cu_3O_{6.6}$ $T_c$-pressure phase diagram by ODMR technology at low temperatures.

**Acknowledgements**

We thank Gang-Qin Liu, En-Ke Liu, and Tao Wu for their helpful discussion. This work was supported by the Innovation Program for Quantum Science and Technology (Grant Nos. 2021ZD0301400, 2021ZD0301200), the National Natural Science Foundation of China (Grant Nos. U19A2075, 11975221, 11874361, 51672279, 51727806, 11774354, 61905233, 61725504, 11804330, and 11821404), the Science Challenge Project (Grant No. TZ2016001), the CAS Innovation Grant (Grant No. CXJJ-19-B08), and the CAS HFIPS Director's Fund (Grant No. YZJJ202102, 2021YZGH03). Anhui Initiative in Quantum Information Technologies (AHY060300) and the Fundamental Research Funds for the Central Universities (Grant No. WK2030380017). X. D. Liu is grateful for the support from the Youth Innovation Promotion Association of CAS (No. 2021446) and Anhui key research and development program(2022h11020007), and J. F. Wang also acknowledges financial support from the Science Specialty Program of Sichuan University (Grant No. 2020SCUNL210). This work was partially carried out at the USTC Center for Micro and Nanoscale Research and Fabrication. We thank Hefei advanced crystal technologies LTD for the sample preparation.

**Author contributions**

J.-F.W., X.-D. L. and J.-S.X. conceived the experiments. J.-F.W. and L.L. built the experimental setup and performed the measurements with the help of X.-D. L., Q.L., J.-Y.Z., J.-M.C., H.-A. X., W. X., J.-W. Y., W.-X.L., Z.-X.H., Z.-H.L., Z.-H.H., and H.-O.L. L. L., J.-F.W., and X.-D.L. prepared the samples in the SiC-based high-pressure chamber. D.-F.Z. prepared the YBCuO sample. Y.W. and W.L. preformed the implantation of the $V_{Si}$ defects. J.-F.W., J.-S.X., L.L. and X.-D.L. performed the data analysis with contributions from all coauthors. J.-F.W., J.-S.X., X.-D.L. and E.G. wrote the manuscript with contributions from all coauthors. J.-S.X., X.-D.L., E.G., C.-F.L. and G.-C.G. supervised the project. All authors contributed to the discussion of the results.

**Competing interests**

The authors declare no competing interests.

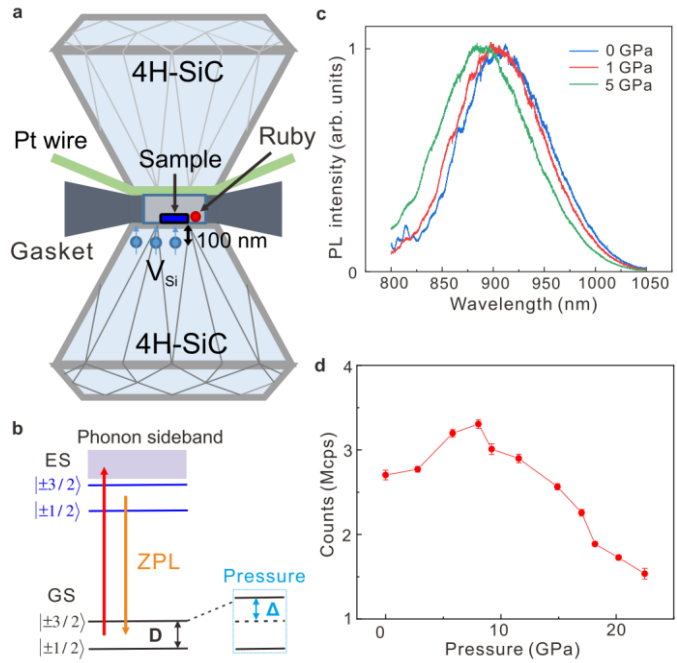

**Fig. 1 The SiC anvil and V$_{Si}$ defect optical properties with pressure. a,** Schematic of a SiC anvil. The samples are placed on the surface of the culet, and local shallow (100 nm) V$_{Si}$ defects are used for *in-situ* magnetic detection. A 10 μm ruby is placed close to the samples to measure the pressure. The c-axis is perpendicular to the culets of the SiC anvil. **b,** Energy levels of V$_{Si}$ defects at high pressure. The red line indicates the 720 nm excitation laser. The pressure changes the ZPL emission and shifts the ground state ZFS parameter $D$ with $\varDelta$. **c,** Room temperature PL spectra of the V$_{Si}$ defects at three representative pressures. **d,** The mean counts of V$_{Si}$ defects with 10 mW laser excitation as a function of the pressure. Error bars are due to the standard deviations of mean counts in an 8×8 μm$^2$ area.

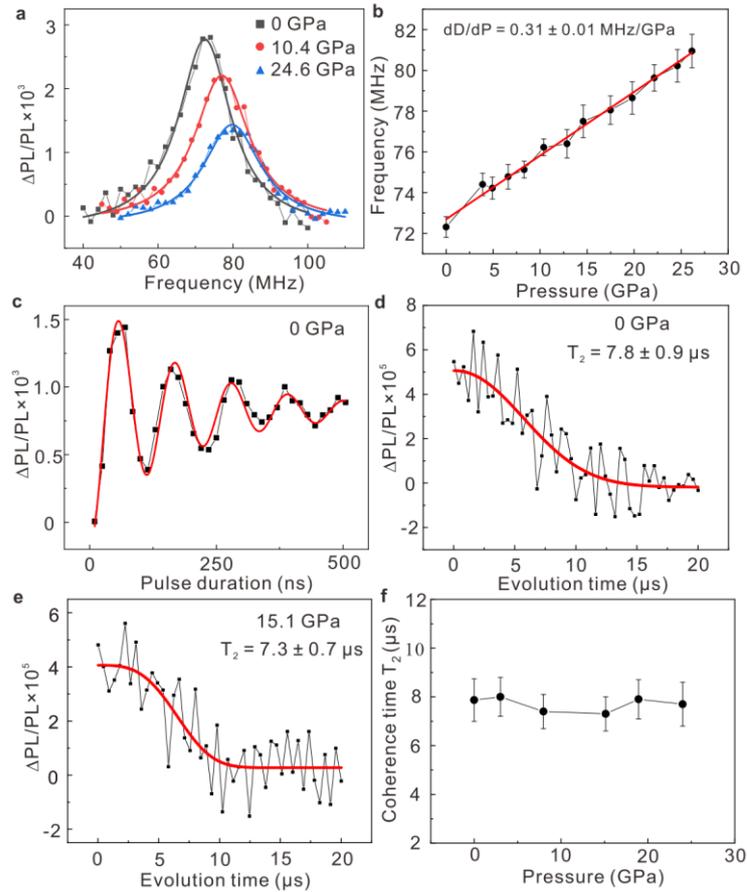

**Fig. 2 The spin properties of V$_{Si}$ defects at high pressures. a,** The ODMR spectra at three different pressures with zero external magnetic field. The solid lines represent Lorentzian fittings. **b,** The mean ZFS parameter D linearly increases with pressure up to 27 GPa. Error bars represent the standard deviations of the mean of the measured D. **c,** Rabi oscillation at 31 Gauss at ambient pressure. The red line is fitted using an exponentially decaying sine function. **d,** and **e,** The spin echo results at ambient pressure and 15.1 GPa, respectively. Red lines represent the exponential decay fittings to the data. **f,** The coherence time $T_2$ as a function of the pressure. Error bars are the data fitting standard deviations of the mean $T_2$.

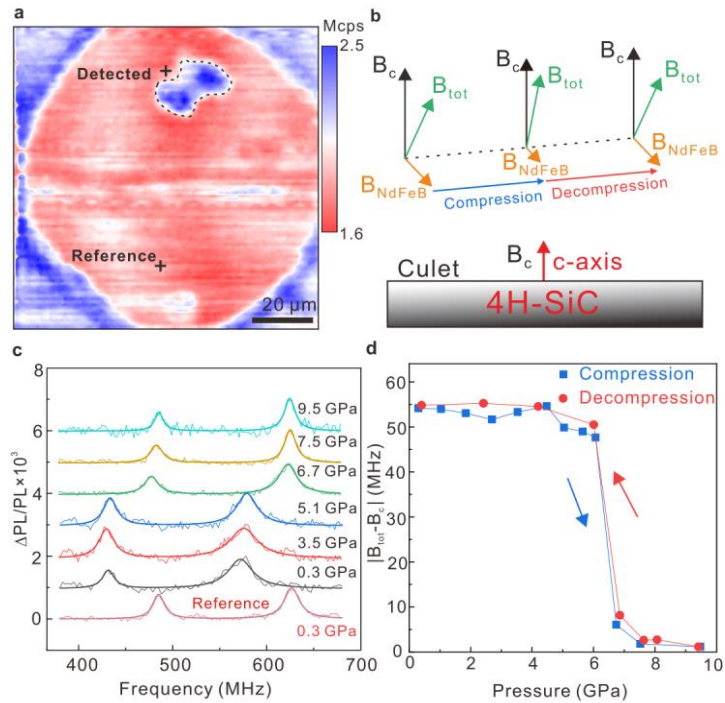

**Fig. 3 The detection of the pressure-induced magnetic phase transition of a Nd$_2$Fe$_{14}$B magnet using shallow V$_{Si}$ defects. a,** The confocal scanning image of the V$_{Si}$ defects and Nd$_2$Fe$_{14}$B sample on the culet surface. The black and red crosses are the detected and reference positions, respectively. The black dashed line region indicates the Nd$_2$Fe$_{14}$B sample. The scale bar is 20 μm. **b,** Three typical local magnetic field vectors during the pressure-induced magnetic phase transition in the compression and decompression processes. The external magnetic field $B_c$ is along the c-axis of the 4H-SiC. The magnetic field of the Nd$_2$Fe$_{14}$B sample and the total magnetic field are labelled $B_{NdFeB}$ and $B_{tot}$, respectively. **c,** The shallow V$_{Si}$ defect ODMR spectra in the detected and reference positions in the culet surface during the compression process. **d,** The inferred magnetic fields of the Nd$_2$Fe$_{14}$B sample were measured by V$_{Si}$ defects during the compression (blue squares) and decompression (red circles) processes.

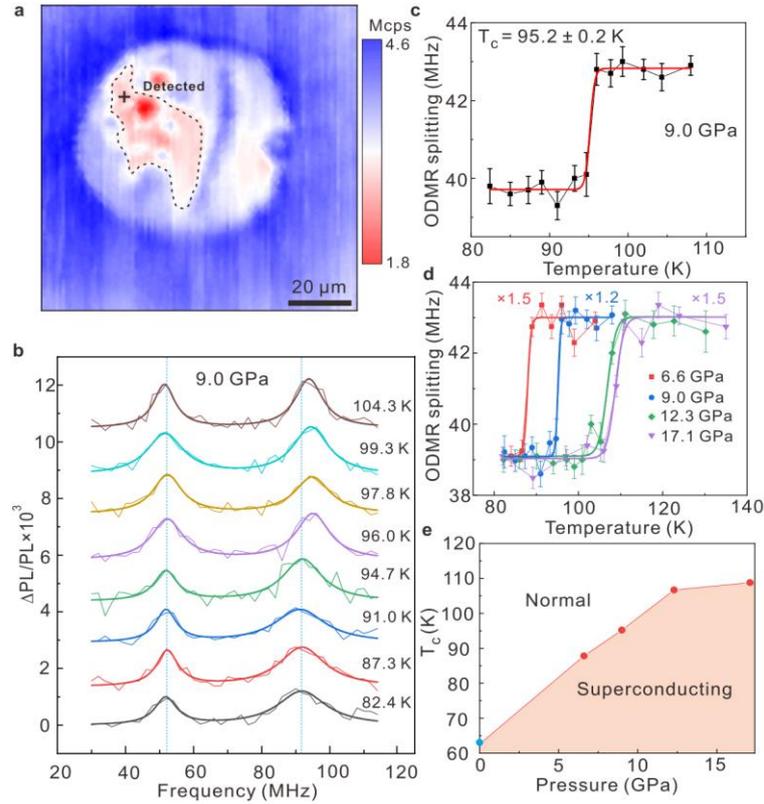

**Fig. 4 Measurement of the temperature–pressure phase diagram of the superconductor YBa$_2$Cu$_3$O$_{6.6}$ using implanted V$_{Si}$ defects. a,** The confocal scanning image of the YBa$_2$Cu$_3$O$_{6.6}$ sample and V$_{Si}$ defects in the culet surface. The black dashed line region marks the investigated YBa$_2$Cu$_3$O$_{6.6}$ sample, and the black cross marks the corresponding detected position. The size scale bar is 20 μm. **b,** The ODMR spectra with superconductivity diamagnetism in the detected position at different temperatures at 9.0 GPa. The dashed lines are guides to an eye only. **c,** The inferred ODMR splitting during the sample superconducting phase transition at 9.0 GPa. The red line is the fitting of the data. The superconducting transition temperature $T_c$ is deduced to be 95.2 ± 0.2 K. **d,** The inferred ODMR splittings as a function of temperature under different pressures. The labelled coefficients are the ODMR splitting magnification times to normalize the data. The error bars in **c** and **d** are the standard deviation of the mean fitted ODMR splittings. **e,** The YBa$_2$Cu$_3$O$_{6.6}$ $T_c$-pressure phase diagram. The $T_c$ under ambient pressure (blue dot) is measured through a magnetic property measurement system (see Supplementary text 2 for more details). The $T_c$ under pressure (red dots) is inferred from the ODMR splittings. The shadow area represents the superconducting state, and the transparent area is the normal state for YBa$_2$Cu$_3$O$_{6.6}$. The error bars obtained from the fitting standard deviations are smaller than the symbol sizes.

**Methods**

**SiC anvil preparation and silicon vacancy generation.** In the experiments, two high-quality single crystal 4H-SiC cubes are used to fabricate SiC anvil with 200 μm diameter culets. As shown in Fig. 1a, high-density shallow $V_{Si}$ defects in a 100 nm depth layer are used for magnetic detection[55]. A nonmagnetic rhenium gasket is used to confine the sample between the two anvils[11]. NaCl is applied as the pressure transmitting medium in all experiments. The *in-situ* pressure is monitored by measuring the PL spectrum of the ruby (approximately 10 μm) in the chamber. A 10 μm platinum wire is placed across the culet surface, which is used for transmitting radiofrequency to control $V_{Si}$ defect spin states. The culet diameter is 200 μm, and the crystalline orientation is the 0001 (c-axis). 20 keV helium ions ($He^+$) with a dose of $1\times10^{13}/cm^2$ are perpendicularly implanted to the culets to generate high-density shallow $V_{Si}$ defects, and the corresponding depth is approximately 100 nm through the stopping and range of ions in matter (SRIM) simulation[55]. We then annealed the SiC anvil at 500 °C for 2 hours to further increase the $V_{Si}$ defect density by ~3 times[55]. The surface $V_{Si}$ density of the defects is estimated to be approximately 7500/μm² (see Supplementary text 1 for more details).

**Experimental setup.** Our setup consists of an in-house built confocal scan microscope equipped with a radiofrequency system[32,55]. Two lasers with wavelengths of 532 nm and 720 nm are used to excite the ruby and $V_{Si}$ defects, respectively. Two 650 nm and 850 nm longpass filters are used for the collection of ruby and $V_{Si}$ defect fluorescence. We adopt a long-working-distance (20 mm) infrared objective (0.4 N.A., Mitutoyo) to excite the samples and collect the fluorescence. A single-photon counting module (SPCM-AQRH-14-FC) is applied to the fluorescence of $V_{Si}$ defects to determine the average photon counts. A liquid nitrogen temperature range optical cryostat (Oxford Instruments) combined with a confocal system is applied in the low-temperature experiments[56]. Standard lock-in technology is used to detect the ODMR and coherence control signals using a photoreceiver (Femto, OE-200-Si)[20,32]. A c-axis (perpendicular to the culet surface) magnetic field $B_c$ is added to adjust the spin state energy levels. To eliminate the heating by the laser and radiofrequency in the measurements, we use a small laser power (8 mW) and radiofrequency power (25 dBm for RdFeB experiments and 15 dBm for YBCO experiments).

**Data availability**

The data that support the findings of this study are presented in the main text and the Supplementary Information, and are available from the corresponding authors upon reasonable request.

**Methods References**